# The K-X-ray intensity ratios as a tool of examination and thickness measurements of coating layers

A.M. Gójska[a], E.A. Miśta-Jakubowska[a], K. Kozioł[a], A. Wasilewski[a], R. Diduszko[b]

[a] National Centre for Nuclear Research, ul. A. Sołtana 7, Otwock, Poland
[b] Institute of Microelectronics and Photonics, Łukasiewicz Research Network, al. Lotników 32/46, Warsaw, Poland.

Abstract: The ED-XRF (Energy-Dispersive X-ray Fluorescence) measurements and the FLUKA simulations have been made to discuss the possibility of recognition of coating layer as well as to its thickness measurement. In this work the $I_{K\alpha}(Cu)/I_{K\alpha}(Ag)$ intensity ratios as well as $I_{K\beta}(Ag)/I_{K\alpha}(Ag)$ and $I_{K\beta}(Cu)/I_{K\alpha}(Cu)$ for copper samples coated with various thickness of sputtered silver have been analyzed. The results show strong dependence of these factors with coating silver layer thickness.

The measurements show the performance of this method in archaeometry. Since the use of non-destructive methods during tests on ancient silver artifacts may not supply to obtain reliable bulk results and should be considered applicable for only surface analyses, the measured intensity ratios can be applied as a tool to estimation of surface silver enrichment thickness.

Keywords: ED-XRF, K-X-ray intensity ratio, FLUKA, surface silver enrichment, thickness determination

## Introduction

The elemental analysis of archaeological finds is one of primary interests of archaeology. In our research, we focused on copper-based alloys, particularly silver-copper alloys, as they have been extensively used in the jewelry and minting industry throughout history [1–4]. Additionally, these alloys find applications in various other fields. Mixtures of silver and copper, as well as pure silver, exhibited antimicrobial properties. While pure copper has been shown to possess antimicrobial properties [5, 6], it is worth noting that not all authors confirm its antimicrobial activity [7]. Despite the intrinsic electrical conductivity of pure copper being similar to silver, its lower resistance to oxidation and the non-conductive nature of its oxides make it less favorable for applications requiring high stability of properties. Therefore, there has been a need to enhance these properties. The utilization of silver-coated copper provides significant benefits in the electronic industry. The silver coating on the surface of copper nanoparticles, creating core-shell coated nanomaterials, not only improves the oxidation resistance of the copper nanoparticles [8] but also preserves their excellent conductivity [9]. As a result, silver-coated copper is commonly used in the production of electronic components such as connectors, contacts, and wires, where it ensures electrical conductivity. The design of modern lithium-ion battery connections for electric vehicles is based on the copper as the primary material for the battery connections due to its high conductivity and mechanical strength [10]. One approach to reducing electrical contact resistance at the junction is the application of a silver coating to enhance conductivity. The sample with a silver-coated copper exhibited the lowest contact resistance and power loss. Thus favorable properties, including good electrical and thermal conductivity, corrosion resistance, and durability, make silver-copper alloys and silver-coated copper suitable for the production of electrical cables, batteries, connectors, and medical devices.

A wide variety of methods which involve spectroscopy techniques can be applied for compositional analyses of archaeological artefacts. The main advantage of the spectroscopic methods is their non-destructive character. Therefore, they are commonly used in art and cultural heritage studies. Both ED-XRF (Energy-Dispersive X-ray Fluorescence) [11–25] and WD-XRF (Wave-Dispersive X-ray Fluorescence) [23] as well as other spectroscopy techniques, such as SEM-EDX (Scanning Electron Microscope with Energy Dispersive X-ray Spectroscopy) [1,13,23,26–29], XRPD (X-ray Powder Diffraction) [13,17,27] and PIXE (Proton Induced X-ray Emission) [12,25,23,30] have been widely applied in archaeometry. The different nature of radiation excitement of these methods results in different depth of interaction. Thus, the informations are collected from various depths.

One of the most widely used materials in the ancient world was a silver-copper alloy. It was mainly used to produce coins and jewelry. In the case of research on the elemental composition of historical/archaeological objects produced based on Ag-Cu alloy [15,31], the problem of surface silver enrichment often appears [15,32–42]. Surface silver enrichment may arise as a result of intentional minting and/or conservation activities involving the extraction of copper from the surface by bathing e.g., in tartaric acid (historical technical technique) or currently in chemical reagents, i.e., sodium edetate: $C_{10}H_{14}N_2O_8Na_2$ and potassium sodium tartrate: $KNaC_4H_4O_6 \cdot 4H_2O$, segregation during casting or annealing and corrosion processes [32,37,40,41].

Last years non-invasive XRF techniques have been used for thickness estimation of coating layers [4,43–47]. Cesareo et al. developed a general method based on ED-XRF analysis to determine the thickness of multi-layered materials [45,46]. The method utilized the intensity ratios of Kα/Kβ or Lα/Lβ X-ray lines for selected elements present in multi-layered objects of various materials containing Au, Ag, Cu, Pb, Zn, Ni in order to study the structure of pre-Columbian artifacts. Next, Lopes et al. [44] used also Partial Least Squares regression (PLS) method for this same purpose. Furthermore, Brocchieri et al. [4,43,47] studied the Kα/Kβ or Lα/Lβ ratios in order to examine the thickness of Ag and Au layers coating Ag, Cu, Fe, and Pb substrates.

Recently, the Monte Carlo techniques have been used for quantitative X-ray fluorescence analysis of various alloys with a covering layer [48–53]. In particular, Giurlani et al. [48] conducted Monte Carlo simulations for Au, Pd, Ni layers of various thicknesses coating Cu substrates in order to construct calibration curves for X-ray intensity ratios. In that work, the authors proposed a method that combines the acquisition of energy dispersive microanalysis spectra with Monte Carlo simulation. In [49, 50], the authors described a semi-quantitative approach to coating thickness measurement based on the construction of calibration curves using simulated XRF spectra generated through Monte Carlo simulations. Pessanha et al. [51] used Monte Carlo simulations for Au Lα/Lβ ratio to determine the thickness of gold foils. Trojek [52] applied the iterative Monte Carlo simulations to evaluation of XRF data obtained for bronze materials covered by Al and Fe foils.

The aim of this work was examination of spectroscopic techniques in study of copper coated with silver layer. The copper sheets with different thickness of silver surface layer emulating the archaeological silver-copper alloys with silver enrichment were made and the $I_{K\alpha}(Cu)/I_{K\alpha}(Ag)$ intensity ratios as well as $I_{K\beta}(Ag)/I_{K\alpha}(Ag)$ and $I_{K\beta}(Cu)/I_{K\alpha}(Cu)$ ones were analysed. The measurement results were applied to the silver thickness determination by use of spectroscopic methods. The results were compared with the FLUKA simulations.

**Materials and Methods**

**Materials**

The research materials have been made in the National Centre for Nuclear Research (NCBJ), Otwock, Poland.

The copper samples were sputtered with silver by use of Quorum Q150T ES Sputter Coater. The diameter of Cu samples was 1.33 cm and the thickness of the silver layer, simulating surface silver enrichment, was 4.9, 10.2, 15.0, 31.1 and 55.9 µm (±3%). The Cu samples were cleaned and then placed on a rotating stage. Due to the rotation, the coating was evenly distributed. The coating process took place in a vacuum of $1*10^{-4}$ mbar.

The study also utilized silver-copper alloys with different compositions. The alloys sourced from *ESPI Metals* included compositions of 90% Ag and 10% Cu, 80% Ag and 20% Cu, and 75% Ag and 25% Cu. Similarly, the alloys obtained from *Goodfellow* consisted of 5% Ag and 95% Cu, as well as 10% Ag and 90% Cu.

**Analytical Methods**

The analyses were performed using ED-XRF technique. In this work we used the EDX 3600H spectrometer equipped in Rh anode. The XRF spectra of studied samples were induced by X-ray tube irradiation on 8 mm diameter surface. The X-ray tube was used with anode voltage and current equal to 40.8 kV and 290 µA. The fluorescence spectra of samples were registered with Si detector positioned 45° to the surface of samples [31]. The energy resolution at 5.9 keV value is 150 ± 5 eV. The distances between the X-ray source and the measured sample (SS) and between the sample and the detector (SD) were set both to 20 mm. The spectra were collected for periods of 300 s. The measured spectra of the samples under study are presented in Fig. 1. The $K\alpha_{1,2}$ (K-$L_{2,3}$ transitions) and the $K\beta_{1,3}+K\beta_2$ (K-$M_{2,3}$ combined with K-$N_{2,3}$ transitions) X-ray lines have been measured.

The FLUKA simulations were applied for generation of X-ray fluorescence spectra for Cu-Ag sandwich: copper layered with different silver thickness [54]. The FLUKA code makes use of the Evaluated Photon Data Library (EPDL97) [55], which provides detailed tables of photon interaction data including photoionization, photoexcitation, coherent and incoherent scattering, as well as pair and triplet production cross sections. This program computes X-ray intensities to simulate real experimental settings, hence the assumption of the simulation reflected the actual measurement conditions (Fig. 2). The experimental setup, replicated in the calculations, includes a model X-ray tube featuring a Rh anode with a thickness of 1 mm and a Be window with a thickness of 1 mm, housed in a tungsten casing with a thickness of 1 mm. The X-ray radiation was generated from the Rh tube through the excitation of a monoenergetic electron beam with a flux exhibiting a flat distribution, $\Phi = 1$ cm. The X-ray tube was configured to operate at 40.8 keV with a current of 290 µA (equivalent to $1.81\times10^{15}$ e/s). In the first stage, $5.9\times10^{11}$ electrons were generated in 2000 parallel processes. On the outer surface of the Be window, $2.94\times10^7$ photons were recorded. These photons serve as the radiation source in the calculations of the second stage. The calculation time lasted for 14 days. In stage 2, calculations were performed on 820 parallel processes for each thickness of the silver layer, with a calculation time of 1.5 days per case. The dose was recorded in a detector (Si) as a single event triggered by each photon separately. In total, 16 TB of raw data was converted into 14.4 GB of useful data.

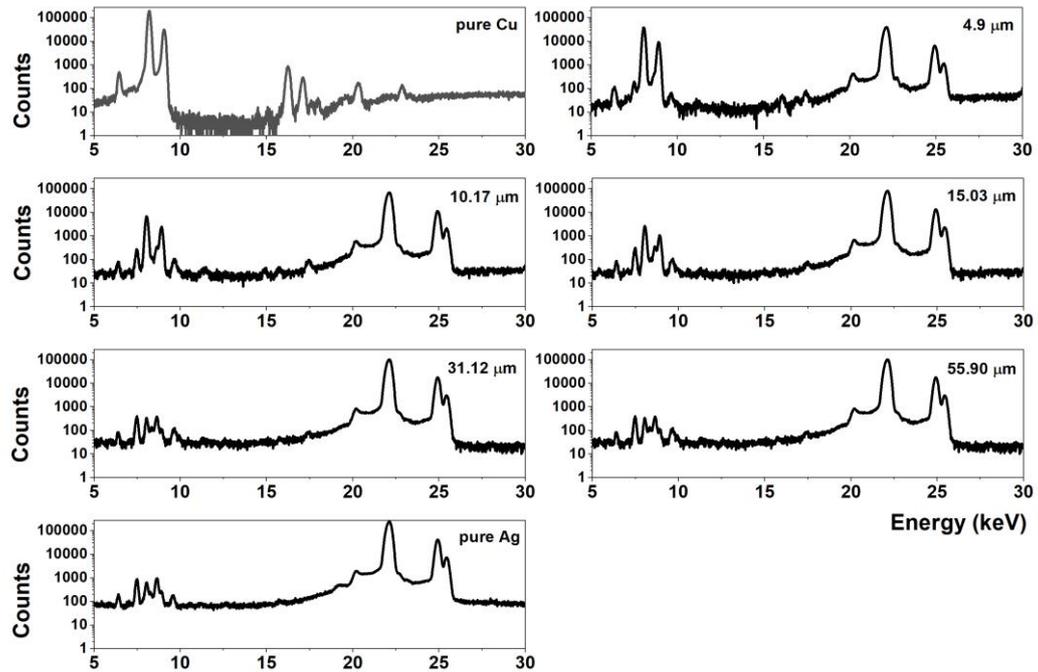

Fig. 1. Measured ED-XRF spectra of copper samples sputtered with different silver thickness.

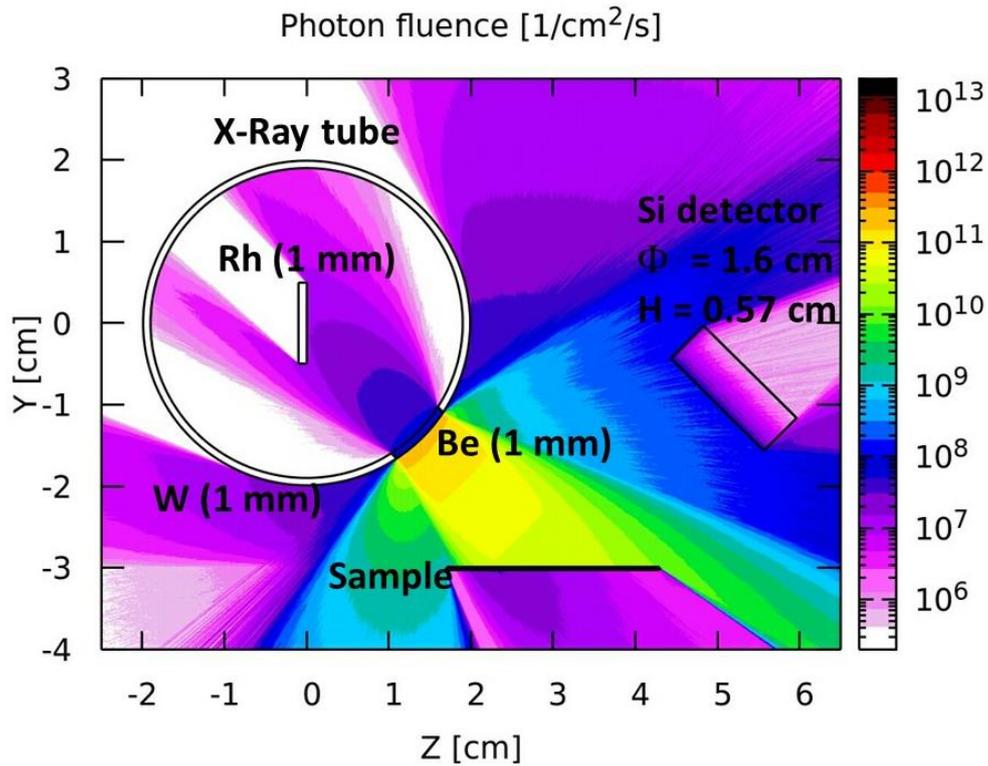

Fig. 2. Experimental setup and photon fluence reproduced in the calculations

**Measurement results**

The presented experimental results were compared with the FLUKA simulations. In Fig. 3. the experimental and simulated $I_{K\alpha}(Cu)/I_{K\alpha}(Ag)$ intensity ratios vs thickness of silver layer is presented. The results prove a decrease of copper signal with an increase in silver layer thickness thus the $I_{K\alpha}(Cu)/I_{K\alpha}(Ag)$ intensity ratio decreases with increasing of surface silver enrichment. The 50 keV X-rays from ED-XRF can interact with about 100 μm deep volumes of the sample. However, the K-X-rays from Cu is carrying only 8.049 keV and cannot escape from this depth of the sample if the Ag surface enrichment layer is thick. It is natural since the copper signal recorded in the detector strongly depends on the thickness of the absorbing layer. As one can see from Fig. 4, the attenuation of $I_{K\alpha}(Cu)$ in silver grows exponentially reaching plateau at a silver layer thickness of about 20 micrometres [56]. Therefore, it is not possible to measure the bulk alloy covered with silver enrichment by use of spectroscopic methods. The calculated attenuation of copper in silver alloy is presented in Fig. 4 [56]. Above 20 μm the signal is strongly attenuated (for surface silver enrichment of 55 μm only 0.02% of Cu-Kα is transmitted). Our results support the findings of Brocchieri et al. [4], where the similar dependence of $I_{K\alpha}(Cu)/I_{K\alpha}(Ag)$ intensity ratio on Ag layer thickness has been studied.

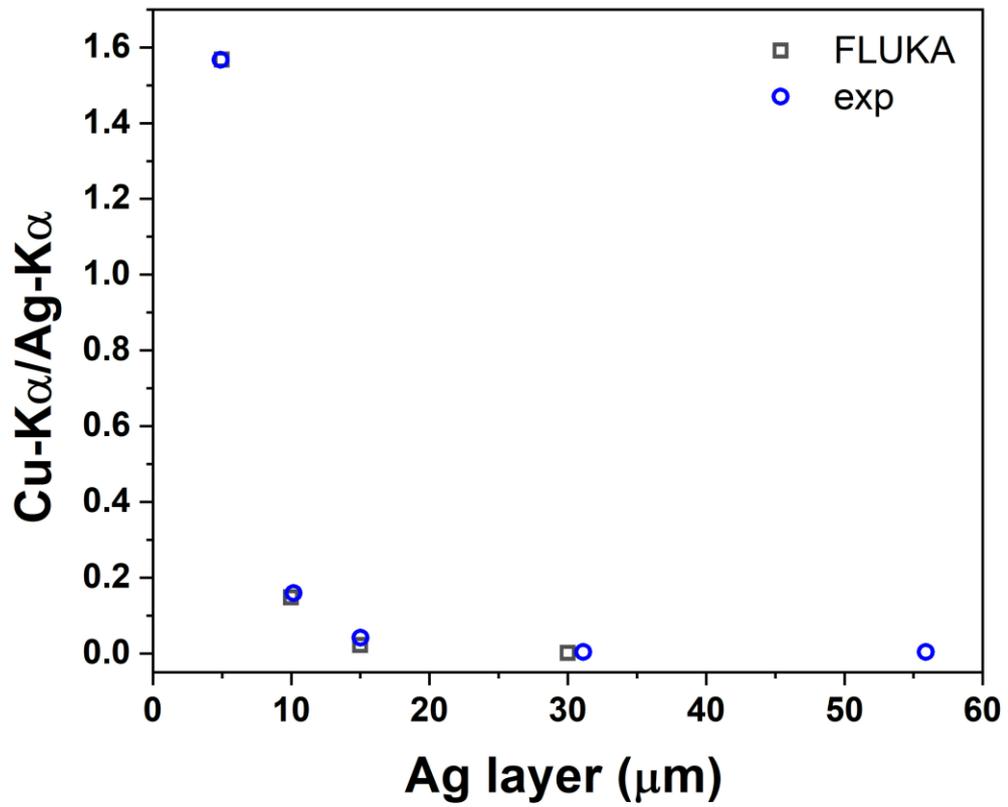

Fig. 3. The $I_{K\alpha}(Cu)/I_{K\alpha}(Ag)$ intensity ratio vs thickness of silver layer on Ag-coated Cu samples seen by use of ED-XRF spectrometer.

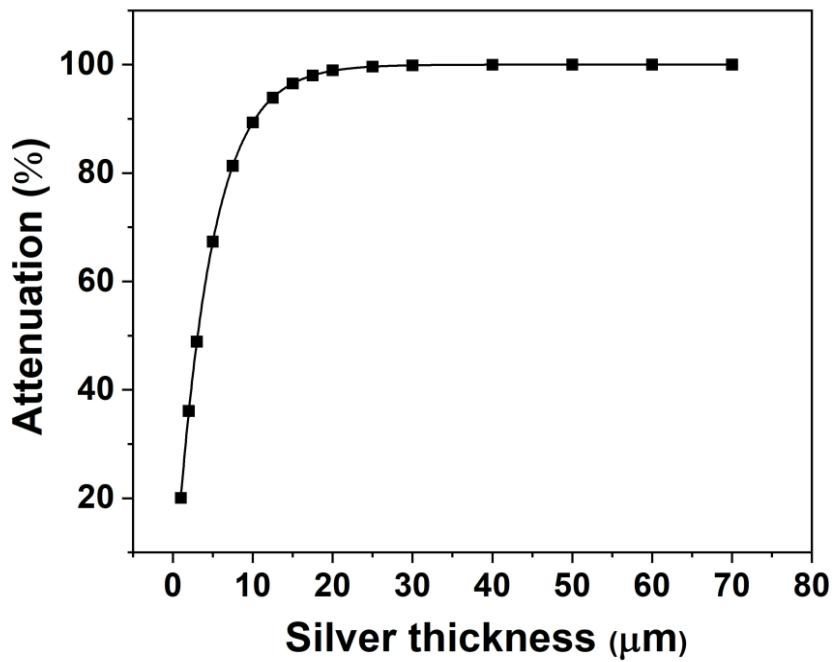

Fig. 4. Cu-Kα (8.049 keV) X-rays attenuation vs silver thickness (calculated with [46]).

The experimental and simulated $I_{K\beta}(Ag)/I_{K\alpha}(Ag)$ intensity ratios vs thickness of silver layer is presented in Fig. 5. As the sputtering increases, the intensity ratio increases towards the value for pure silver. In the case of surface silver enrichment, the $I_{K\beta}(Ag)/I_{K\alpha}(Ag)$ ratio is lower than the one for pure silver. An opposite behavior is seen in the case of Ag-Cu alloys, when the $I_{K\beta}(Ag)/I_{K\alpha}(Ag)$ ratio is higher than the one for pure silver and it increases with decreasing silver content in the alloy (see Fig. 6). Therefore, these types of measurements are a valuable guide in identifying materials supposed to be a mix of Ag and Cu – if the $I_{K\beta}(Ag)/I_{K\alpha}(Ag)$ ratio is lower than the one for pure silver it means the material is a layered material, otherwise it is an alloy. The $I_{K\beta}(Ag)/I_{K\alpha}(Ag)$ intensity ratio vs thickness of silver layer dependence has been studied before by Brocchieri et al. [4] for thickness range between 0.2 and 15.7 μm. However we have measured $K\beta_{1,3}+K\beta_2$ combined peak, while Brocchieri et al. measured likely only $K\beta_{1,3}$ peak. So, on Fig. 5 we presented also our numbers for $K\beta_{1,3}$ peak only. As one can see from this figure, in the case of $K\beta_{1,3}$ lines our number are similar to Brocchieri et al. numbers in the range of 5-15 μm. In the work of Brocchieri et al. the fit function to the experimental $K\alpha/K\beta_{1,3}$ ratio has been presented. As one can see from Fig. 5, this Brocchieri et al. fit function is not correct for silver thickness larger than about 15 μm.

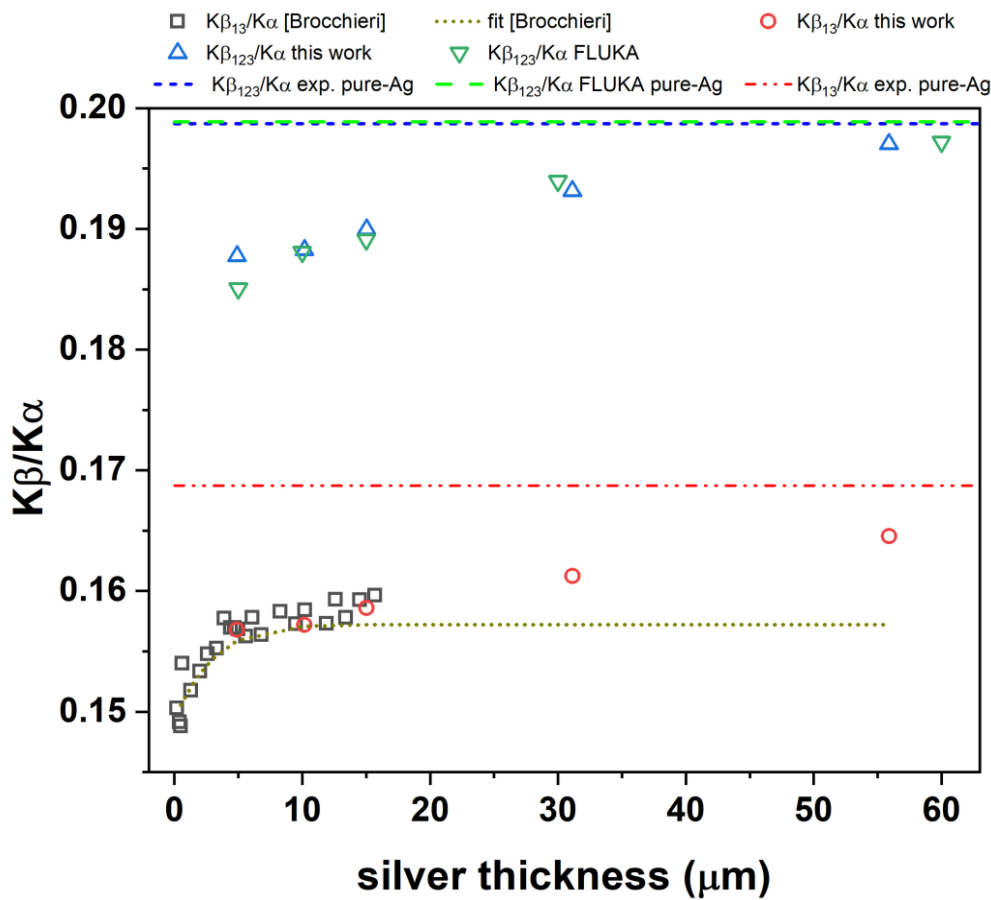

Fig. 5. The $I_{K\beta}(Ag)/I_{K\alpha}(Ag)$ intensity ratio vs thickness of silver layer on Ag-coated Cu samples.

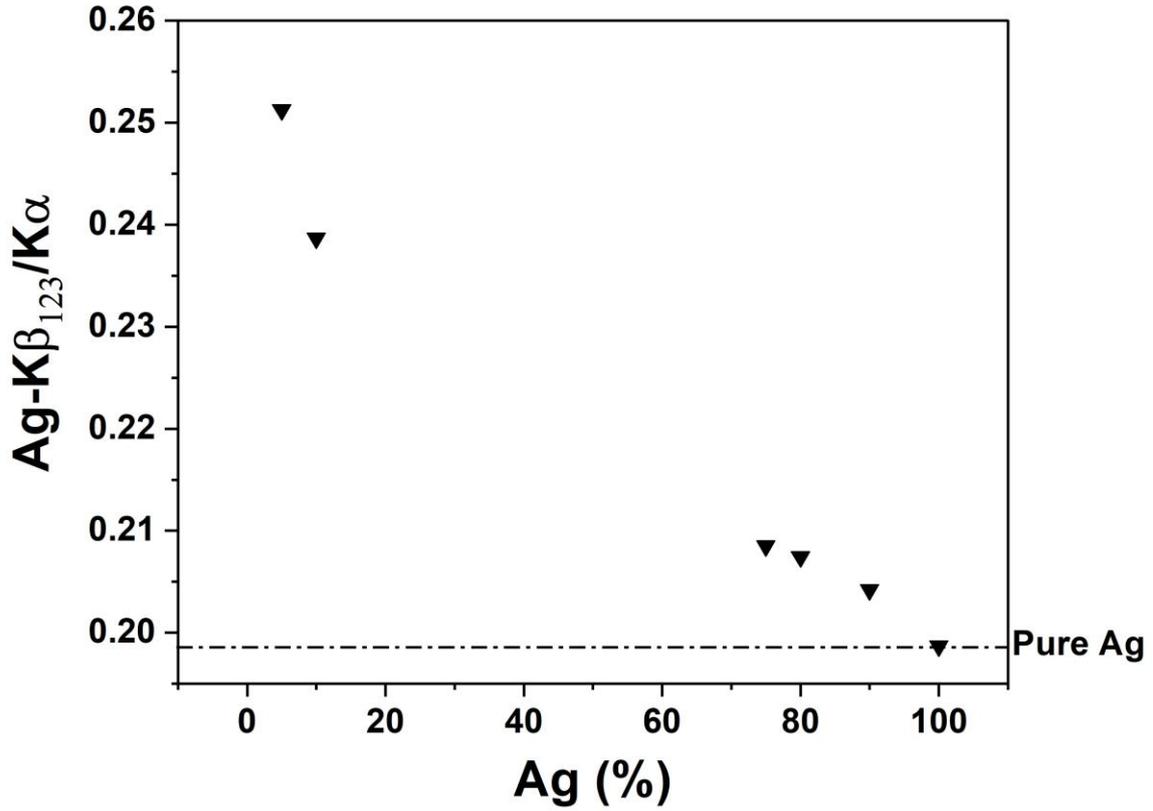

Fig. 6. The measured $I_{K\beta}(Ag)/I_{K\alpha}(Ag)$ intensity ratio vs silver content in bulk samples of Ag-Cu alloy.

The measuring of the Kα and Kβ intensities of base material can be employed to determination of coating thickness [44,57–59]. The ratio of base material is expressed by

$$\frac{I_{K\alpha}}{I_{K\beta}} = \frac{I_{K\alpha(0)}}{I_{K\beta(0)}} exp[(\mu_{K\alpha} - \mu_{K\beta})x]$$

where

$\mu_{K\alpha}$ and $\mu_{K\beta}$ are the attenuation coefficients of silver at energy of Cu-Kα (8.048 keV) and Cu-Kβ (8.905 keV) lines, respectively, $\alpha$ and $\frac{I_{K\alpha(0)}}{I_{K\beta(0)}}$ are the K-X-ray intensity ratios of copper with and without coating, and $x$ is the silver layer thickness.

Thus, coating thickness can be estimated using the following:

$$x = \frac{1}{\mu_{K\alpha} - \mu_{K\beta}} ln \frac{\frac{I_{K\alpha}}{I_{K\beta}}}{\frac{I_{K\alpha(0)}}{I_{K\beta(0)}}},$$

The comparison between XRF estimated and nominal silver layer thickness is presented in Fig. 7. The values estimated by use of peak intensities are in good agreement with coating of nominal 4.9 µm and 10.2 µm layer thickness and difference is less than 1%. The discrepancy between estimated and nominal values for 15.0 µm thickness is 11%. For coating layer thickness above 20 µm this method cannot be use due to strong attenuation (99%) of Cu signal. The almost complete absorption causes a disturbed and not reliable signal of the copper peaks (see Fig. 8).

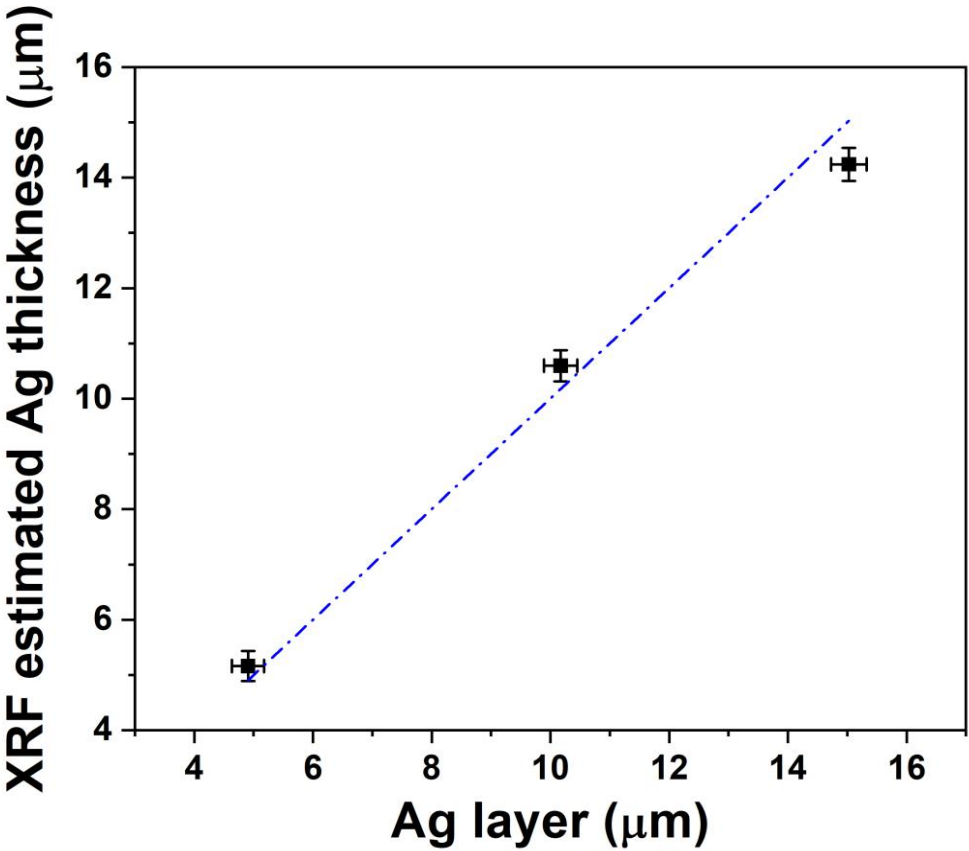

Fig. 7. XRF estimated silver layer thickness vs nominal silver layer thickness on Ag-coated Cu samples. The blue line is an ideal one-to-one linearity between XRF estimated and nominal silver layer thicknesses.

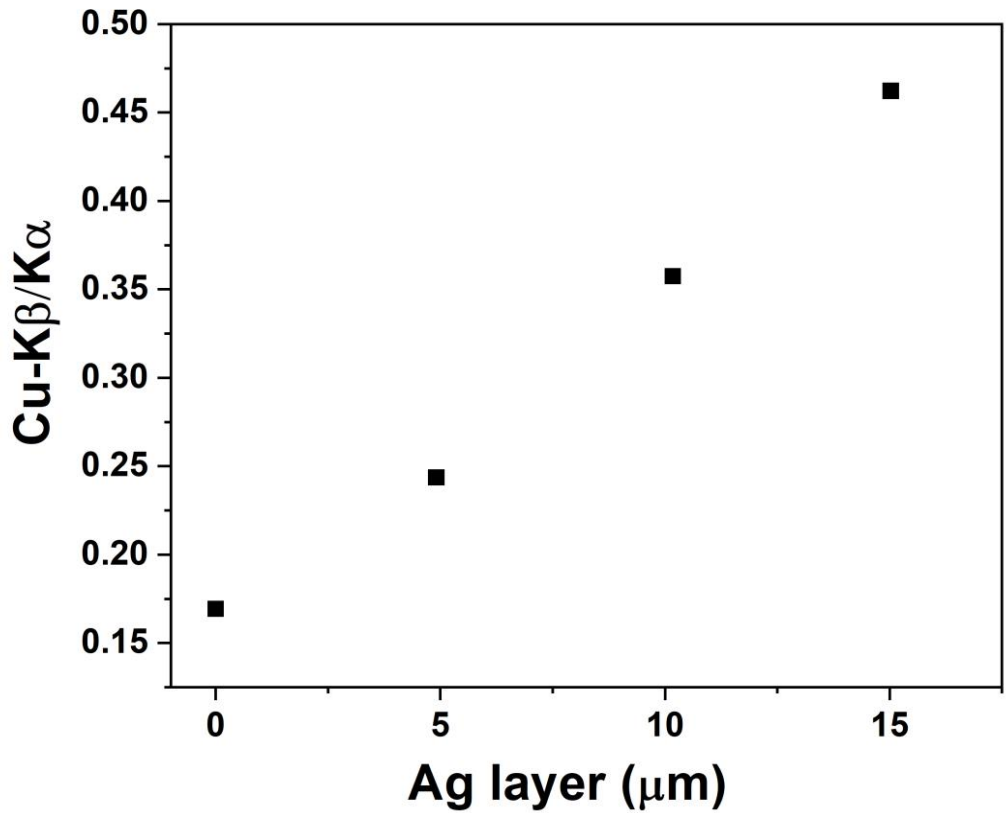

Fig. 8. The $I_{K\beta}(Cu)/I_{K\alpha}(Cu)$ intensity ratio vs thickness of silver layer on Ag-coated Cu samples.

**Conclusions**

The ED-XRF measurements and FLUKA simulations have been made in order to discuss the use of X-ray fluorescence as a tool of examination of coating layers. The spectra of copper samples sputtered with silver thickness of 4.9, 10.2, 15.0, 31.1 and 55.9 μm were registered. The $I_{K\beta}(Ag)/I_{K\alpha}(Ag)$ intensity ratios (measured and simulated) for Cu samples layered with different thickness of Ag while measured $I_{K\beta}(Ag)/I_{K\alpha}(Ag)$ intensity ratios for different Ag-Cu alloys have been analyzed. As the sputtering increases, the intensity ratio increases towards the value for pure silver. In the case of coating layer the $I_{K\beta}(Ag)/I_{K\alpha}(Ag)$ ratio is lower than the one for pure silver. An opposite behaviour is observed in the case of Ag-Cu alloys, when the $I_{K\beta}(Ag)/I_{K\alpha}(Ag)$ ratio is higher than the one for pure silver and it increases with decreasing silver content in the alloy. Therefore, these types of measurements are a valuable guide in identifying materials supposed to be a mix of Ag and Cu – if the $I_{K\beta}(Ag)/I_{K\alpha}(Ag)$ ratio is lower than the one for pure silver it means the material is a layered material, otherwise it is an alloy.

Since the X-ray fluorescence is commonly used spectroscopic technique in archaeometry the K-X-ray intensity ratio can be an indicator for surface silver enrichment. The most critical issue is that the information depth is too low for a precise validation of the Cu content using non-invasive methods such as ED-XRF. Although X-ray radiation (XRF) with energy of about 40.9 keV penetrates the material up to approximately 100 microns the feedback signal registered by the detector comes from a depth of few tens to few microns. The similar problems appear for other spectroscopic techniques, e.g., PIXE or SEM-EDX. Although protons (PIXE) with energy of 2.75-3.00 MeV penetrate up to 32.13-36.74 μm [56], only the elemental composition of near surface layers can be obtained. In the case of SEM-EDX only ~2 μm surface thickness can be studied. Therefore, results of spectroscopic techniques can be used for preliminary analyses of the elemental composition. In order to obtain more reliable results, measurement should be done on the cut, which is practically impossible without the destruction or serious damage of the artefact. Although the use of non-destructive methods during tests on ancient silver artefacts should be considered applicable for only surface analyses, the measured of K-X-ray intensity ratios can be applied as a tool of estimation of surface silver enrichment thickness. The measuring of Ag-K-X-ray intensity ratio can show whether we are dealing with a different composition [60]: higher K-X-ray intensity ratio in relation to pure Ag or with silver enrichment: lower K-X-ray intensity ratio in relation to pure silver. Additionally, the measurement of base- K-X-ray intensity ratio can be used as a tool to decide the unknown coating thickness what is importance from archaeometric point of view.

### Acknowledgments

The author would like to thank Jason Welch for language proofreading.